# Mining Frequent Itemsets over Uncertain Databases

Yongxin Tong †   Lei Chen †   Yurong Cheng ‡   Philip S. Yu §
†Hong Kong University of Science & Technology, Hong Kong, China
‡Northeastern University, China
§University of Illinois at Chicago, USA

†{yxtong, leichen}@cse.ust.hk,   ‡cyrneu@gmail.com,   §psyu@cs.uic.edu

## ABSTRACT

In recent years, due to the wide applications of uncertain data, mining frequent itemsets over uncertain databases has attracted much attention. In uncertain databases, the support of an itemset is a random variable instead of a fixed occurrence counting of this itemset. Thus, unlike the corresponding problem in deterministic databases where the frequent itemset has a unique definition, the frequent itemset under uncertain environments has two different definitions so far. The first definition, referred as the expected support-based frequent itemset, employs the expectation of the support of an itemset to measure whether this itemset is frequent. The second definition, referred as the probabilistic frequent itemset, uses the probability of the support of an itemset to measure its frequency. Thus, existing work on mining frequent itemsets over uncertain databases is divided into two different groups and no study is conducted to comprehensively compare the two different definitions. In addition, since no uniform experimental platform exists, current solutions for the same definition even generate inconsistent results. In this paper, we firstly aim to clarify the relationship between the two different definitions. Through extensive experiments, we verify that the two definitions have a tight connection and can be unified together when the size of data is large enough. Secondly, we provide baseline implementations of eight existing representative algorithms and test their performances with uniform measures fairly. Finally, according to the fair tests over many different benchmark data sets, we clarify several existing inconsistent conclusions and discuss some new findings.

## 1. INTRODUCTION

Recently, with many new applications, such as sensor network monitoring [23, 24, 26], moving object search [13, 14, 15] and protein-protein interaction (PPI) network analysis [29], uncertain data mining has become a hot topic in data mining communities [3, 4, 5, 6, 20, 21]. Since the problem of frequent itemset mining is fundamental in data mining area, mining frequent itemsets over uncertain databases has also attracted much attention [4, 9, 10, 11, 17, 18, 22, 28, 30, 31, 33]. For example, with the popularization of wireless sensor networks, wireless sensor network systems collect huge amount of data. However, due to the inherent uncertainty of sensors, the collected data are often inaccurate. For the probability-included uncertain data, how can we discover frequent patterns (itemsets) so that the users can understand the hidden rules in data? The inherent probability property of data is ignored if we simply apply the traditional method of frequent itemset mining in deterministic data to uncertain data. Thus, it is necessary to design specialized algorithms for mining frequent itemsets over uncertain databases.

Before finding frequent itemsets over uncertain databases, the definition of the frequent itemset is the most essential issue. In deterministic data, it is clear that an itemset is frequent if and only if the support (frequency) of such itemset is not smaller than a specified minimum support, $min\_sup$ [7, 8, 19, 32]. However, different from the deterministic case, the definition of a frequent itemset over uncertain data has two different semantic explanations: *expected support-based frequent itemset* [4, 18] and *probabilistic frequent itemset* [9]. Both of which consider the support of an itemset as a discrete random variable. However, the two definitions are different on using the random variable to define frequent itemsets. In the definition of the expected support-based frequent itemset, the expectation of the support of an itemset is defined as the measurement, called as the expected support of this itemset. In this definition [4, 17, 18, 22], *an itemset is frequent if and only if the expected support of such itemset is no less than a specified minimum expected support threshold, $min\_esup$*. In the definition of probabilistic frequent itemset [9, 28, 31], the probability that an itemset appears at least the minimum support ($min\_sup$) times is defined as the measurement, called as the frequent probability of an itemset, and *an itemset is frequent if and only if the frequent probability of such itemset is larger than a given probabilistic threshold*.

The definition of expected support-based frequent itemset uses the expectation to measure the uncertainty, which is a simply extension of the definition of the frequent itemset in deterministic data. The definition of probabilistic frequent itemset includes the complete probability distribution of the support of an itemset. Although the expectation is known as an important statistic, it cannot show the complete probability distribution. Most prior researches believe that the two definitions should be studied respectively [9, 28, 31].

However, we find that the two definitions have a rather close connection. Both definitions consider the support of an itemset as a random variable following the Poisson Binomial distribution [2], that is the expected support of an itemset equals to the expectation of the random variable. Consequently, computing the frequent probability of an itemset is equivalent to calculating the cumulative distribution function of this random variable. In addition, the existing mathematical theory shows that Poisson distribution and Normal distribution can approximate Poisson Binomial distribution under high confidence [31, 10]. Based on the Lyapunov Central Limit Theory [25], the Normal distribution converges to Poisson Binomial distribution with high probability. Moreover, the Poisson Binomial distribution has a sound property: the computation of the expectation and variance are the same in terms of computational complexity. Therefore, the frequent probability of an itemset can be directly computed as long as we know the expected value and variance of the support of such itemset when the number of transactions in the uncertain database is large enough [10] (due to the requirement of the Lyapunov Central Limit Theory). In other words, the second definition is identical to the first definition if the first definition also considers the variance of the support at the same time. Moreover, another interesting result is that existing algorithms for mining expected support-based frequent itemsets are applicable to the problem of mining probabilistic frequent itemsets as long as they also calculate the variance of the support of each itemset when they calculate each expected support. Thus, the efficiency of mining probabilistic frequent itemsets can be greatly improved due to the existence of many efficient expected support-based frequent itemset mining algorithms. In this paper, we verify the conclusion through extensive experimental comparisons.

Besides the overlooking of the hidden relationship between the two above definitions, existing research on the same definition also shows contradictory conclusions. For example, in the research of mining expected support-based frequent itemsets, [22] shows that UFP-growth algorithm always outperforms UApriori algorithm with respect to the running time. However, [4] reports that UFP-growth algorithm is always slower than UApriori algorithm. These inconsistent conclusions make later researchers confused about which result is correct.

The lacking of uniform baseline implementations is one of the factors causing the inconsistent conclusions. Therefore, different experimental results originate from discrepancy among many implementation skills, blurring what are the contributions of the algorithms. For instance, the implementation for UFP-growth algorithm uses the "float type" to store each probability. While the implementation for UH-Mine algorithm adopts the "double type". The difference of their memory cost cannot reflect the effectiveness of the two algorithms objectively. Thus, uniform baseline implementations can eliminate interferences from implementation details and report true contributions of each algorithm.

Except uniform baseline implementations, the selection of objective and scientific measures is also one of the most important factors in the fair experimental comparison. Because uncertain data mining algorithms need to process a large amount of data, the running time, memory cost and scalability are basic measures when the correctness of algorithms is guaranteed. In addition, to trade off the accuracy for efficiency, approximate probabilistic frequent itemset mining algorithms are also proposed [10, 31]. For comparing the relationship between the two frequent itemset definitions, we use precision and recall as measures to evaluate the approximation effectiveness. Moreover, since the above inconsistent conclusions may be caused by the dependence on datasets, in this work, we choose six different datasets, three dense ones and three sparse ones with different probability distributions (e.g. Normal distribution Vs. Zipf distribution or High probability Vs. Low probability).

To sum up, we try to achieve the following goals:
- Clarify the relationship of the existing two definitions of frequent itemsets over uncertain databases. In fact, there is a mathematical correlation between them. Thus, the two definitions can be integrated together. Based on this relationship, instead of spending expensive computation cost to mine probabilistic frequent itemsets, we can directly use the solutions for mining expected support-based itemsets as long as the size of data is large enough.
- Verify the contradictory conclusions in the existing research and summarize a series of fair results.
- Provide uniform baseline implementations for all existing representative algorithms under two definitions. These implementations adopt common basic operations and offer a base for comparing with the future work in this area. In addition, we also proposed a novel approximate probabilistic frequent itemset mining algorithm, NDUH-Mine which is combined with two existing classic algorithm: UH-Mine algorithm and Normal distribution-based frequent itemset mining algorithm.
- Propose an objective and sufficient experimental evaluation and test the performances of the existing representative algorithms over extensive benchmarks.

The rest of the paper is organized as follows. In Section 2, we give some basic definitions about mining frequent itemset over uncertain databases. Eight representative algorithms are reviewed in Section 3. Section 4 presents all the experimental comparisons and the performance evaluations. We conclude in Section 5.

## 2. DEFINITIONS

In this section, we give several basic definitions about mining frequent itemsets over uncertain databases.

Let $I = \{i_1, i_2, \ldots, i_n\}$ be a set of distinct items. We name a non-empty subset, $X$, of $I$ as an itemset. For brevity, we use $X = x_1 x_2 \ldots x_n$ to denote itemset $X = \{x_1, x_2, \ldots, x_n\}$. $X$ is a $l-itemset$ if it has $l$ items. Given an uncertain transaction database $UDB$, each transaction is denoted as a tuple $< tid, Y >$ where $tid$ is the transaction identifier, and $Y = \{y_1(p_1), y_2(p_2), \ldots, y_m(p_m)\}$. $Y$ contains m units. Each unit has an item $y_i$ and a probability, $p_i$, denoting the possibility of item $y_i$ appearing in the $tid$ tuple. The number of transactions containing $X$ in $UDB$ is a random variable, denoted as $sup(X)$. Given $UDB$, the expected support-based frequent itemset and probabilistic frequent itemsets are defined as follows.

*Definition 1.* (Expected Support) Given an uncertain transaction database $UDB$ which includes $N$ transactions, and an itemset $X$, the expected support of $X$ is:

$$esup(X) = \sum_{i=1}^{N} p_i(X)$$

Table 1: An Uncertain Database

| TID | Transactions |
|---|---|
| T1 | A (0.8)  B (0.2)  C (0.9)  D (0.7)  F(0.8) |
| T2 | A (0.8)  B (0.7)  C (0.9)  E (0.5) |
| T3 | A (0.5)  C (0.8)  E (0.8)  F (0.3) |
| T4 | B (0.5)  D (0.5)  F (0.7) |

Table 2: The Probability Distribution of sup($A$)

| sup(A) | 0 | 1 | 2 | 3 |
|---|---|---|---|---|
| Probability | 0.1 | 0.18 | 0.4 | 0.32 |

*Definition 2.* (Expected-Support-based Frequent Itemset) Given an uncertain transaction database $UDB$ which includes $N$ transactions, and a minimum expected support ratio, $min\_esup$, an itemset $X$ is an expected support-based frequent itemset if and only if $esup(X) \geq N \times min\_esup$

*Example 1.* (Expected Support-based Frequent Itemset) Given an uncertain database in Table 1 and the minimum expected support, $min\_esup$=0.5, there are only two expected support-based frequent itemsets: $A(2.1)$ and $C(2.6)$ where the number in each bracket is the expected support of the corresponding itemset.

*Definition 3.* (Frequent Probability) Given an uncertain transaction database $UDB$ which includes $N$ transactions, a minimum support ratio $min\_sup$, and an itemset $X$, $X$'s frequent probability, denoted as $Pr(X)$, is shown as follows:

$$Pr(X) = Pr\{sup(X) \geq N \times min\_sup\}$$

*Definition 4.* (Probabilistic Frequent Itemset) Given an uncertain transaction database $UDB$ which includes $N$ transactions, a minimum support ratio $min\_sup$, and a probabilistic frequent threshold $pft$, an itemset $X$ is a probabilistic frequent itemset if $X$'s frequent probability is larger than the probabilistic frequent threshold, namely,

$$Pr(X) = Pr\{sup(X) \geq N \times min\_sup\} > pft$$

*Example 2.* (Probabilistic Frequent Itemset) Given an uncertain database in Table 2, $min\_sup$=0.5, and $pft = 0.7$, the probability distribution of the support of $A$ is shown in Table 2. So, the frequent probability of $A$ is: $Pr(X) = Pr\{sup(A) \geq 4 \times 0.5\} = Pr\{sup(A) \geq 2\} = Pr\{sup(A) = 2\} + Pr\{sup(A) = 3\} = 0.4 + 0.32 > 0.7 = pft$. Thus, $\{A\}$ is a probabilistic frequent itemset.

## 3. ALGORITHMS OF FREQUENT ITEMSET MINING

We categorize the eight representative algorithms into three groups. The first group is the expected support-based frequent algorithms. These algorithms aim to find all expected support-based frequent itemsets. For each itemset, these algorithms only consider the expected support to measure its frequency. The complexity of computing the expected support of an itemset is O($N$), where $N$ is the number of transactions. The second group is the exact probabilistic frequent algorithms. These algorithms discover all probabilistic frequent itemsets and report exact frequent probability for each itemset. Due to complexity of computing the exact frequent probability instead of the simple expectation, these algorithms need to spend at least O($NlogN$) computation cost for each itemset. Moreover, in order to avoid redundant processing, the Chernoff bound-based pruning is a way to reduce the running time of this group of algorithms. The third group is the approximate probabilistic frequent algorithms. Due to the sound properties of the Poisson Binomial distribution, this group of algorithms can obtain the approximate frequent probability with high quality by only acquiring the first moment (*expectation*) and the second moment (*variance*). Therefore, the third kind of algorithms have the O($N$) computation cost and return the complete probability information when uncertain databases are large enough. To sum up, the third kind of algorithms actually build a bridge between two different definitions of frequent itemsets over uncertain databases.

### 3.1 Expected Support-based Frequent Algorithms

In this subsection, we summarize three the most representative expected support-based frequent itemset mining algorithms: $UApriori$ [17, 18], $UFP - growth$ [22], $UH - Mine$ [4]. The first algorithm is based on the generate-and-test framework employing the breath-first search strategy. The other two algorithms are based on the divide-and-conquer framework which uses the depth-first search strategy. Although Apriori algorithm is slower than the other two algorithms in deterministic databases, UApriori which is the uncertain version of Apriori, actually performs rather well among the three algorithms and is usually the fastest one in dense uncertain datasets based on our experimental results in Section 4. We further explain three algorithms in the following subsections and Section 4.

#### 3.1.1 UApriori

The first expected support-based frequent itemset mining algorithm was proposed by Chui et al. in 2007 [18]. This algorithm extends the well-known Apriori algorithm [17, 18] to the uncertain environment and uses the generate-and-test framework to find all expected support-based frequent itemsets. We generally introduced UApriori algorithm as follows. The algorithm first finds all the expected support-based frequent items firstly. Then, it repeatedly joins all expected support-based frequent $i$-itemsets to produce $i+1$-itemset candidates and test $i+1$-itemset candidates to obtain expected support-based frequent $i+1$-itemsets. Finally, it ends when no expected support-based frequent $i+1$-itemsets are generated.

Fortunately, the well-known downward closure property [8] still works in uncertain databases. Thus, the traditional Apriori pruning can be used when we check whether an itemset is an expected support-based frequent itemset. In other words, all supersets of this itemset must not be expected support-based frequent itemsets. In addition, several decremental pruning methods [17, 18] were proposed for further improving the efficiency. These methods mainly aim to find the upper bound of the expected support of an itemset as early as possible. Once the upper bound is lower than the minimum expected support, the traditional Apriori pruning can be used. However, the decremental pruning methods depend on the structure of datasets, thus, the most important pruning method in UApriori is still the traditional Apriori pruning.

### 3.1.2 UFP-Growth

UFP-growth algorithm [22] was extended from the FP-growth algorithm [19] which is one of the most well-known pattern mining algorithms in deterministic databases. Similar to the traditional FP-growth algorithm, UFP-growth algorithm also firstly builds an index tree, called UFP-tree to store all information of the uncertain database. Then, based on the UFP-tree, the algorithm recursively builds conditional subtrees and finds expected support-based frequent itemsets. The UFP-tree for the UDB in Table 1 is shown in Figure 1 when $min\_esup$=0.25.

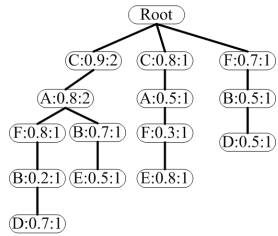

**Figure 1: UFP-Tree**

In the process of building the UFP-tree, it first finds all expected support-based frequent items and orders these items by their expected supports. For the uncertain database in Figure 1, the ordered item list is {$C$:2.6, $A$:2.1, $F$:1.8, $B$:1.4, $E$:1.3, $D$:1.2} where the real number following the colon is the expected support for each item. Based on the list, the algorithm sorts each transaction and inserts the transaction into the UFP-tree. Each node includes three values in UFP-tree. The first value is the label of the item; the second value is the appearance probability of this item; and the third value is the numbers that this node is shared from root to it.

Different from the traditional FP-tree, the compression of UFP-tree is substantially reduced because it is hard to take the advantage of the shared prefix path in the FP-tree under uncertain databases. In the UFP-tree, items may share one node only when their labels and appearance probabilities are both same. Otherwise, items must be presented in two different nodes. In fact, the probabilities in an uncertain database make the corresponding deterministic database become sparse due to fewer shared nodes and paths. Thus, uncertain databases are often considered as the sparse databases. Given an UDB, we have to build many conditional subtrees in the corresponding UFP-tree, which leads much redundant computation. That is also the reason why UFP-growth cannot achieve the similar performance as FP-growth does.

### 3.1.3 UH-Mine

UH-Mine [4] is also based on the divide-and-conquer framework and the depth-first search strategy. The algorithm was extended from the H-Mine algorithm [27] which is classical algorithm in deterministic frequent itemset mining. In particular, H-Mine is quite suitable for sparse databases.

UH-Mine algorithm can be outlined as follows. Firstly, it scans the uncertain database and finds all expected support-based frequent items. Then, the algorithm builds a head table which contains all expected support-based frequent items. For each item, the head table stores three elements: the label of this item, the expected support of such item, and a pointer domain. After building the head table, the algorithm inserts all transactions into the data structure, UH-Struct.

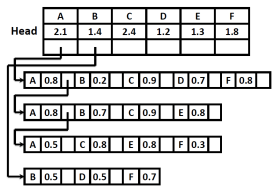

**Figure 2: UH-Struct Generated from Table 1**

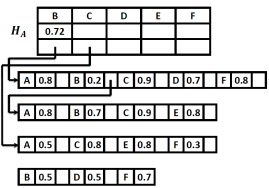

**Figure 3: UH-Struct of Head Table of A**

In this data structure, each item is assigned with its label, its appearing probability and a pointer. The UH-Struct of Table 1 is shown in Figure 2. After building the global UH-Strut, the algorithm uses the depth-first strategy to build the head table in Figure 3 where A is the prefix. Then, the algorithm recursively builds the head tables where different itemsets are prefix and generates all expected support-based frequent itemsets.

In frequent itemset mining over deterministic databases, H-Mine algorithm fails to compress the data structure and use the dynamic frequency order sub-structure, such as conditional FP-trees, which only builds the head tables of different itemsets recursively. Therefore, for the dense databases, the FP-growth is superior because a larger number of items are stored in fewer shared paths. However, for the sparse databases, H-Mine is faster when building head tables of all levels is faster than building all conditional subtrees. Thus, it is quite likely that H-Mine is better than FP-growth in sparse databases. As discussed in Section 3.1.2, uncertain databases are quite sparse databases, so UH-Mine extended from H-Mine always has the good performance.

**Comparison of Three Algorithm Frameworks:** Moreover, The search strategies and data structures of three above algorithms are shown in Table 3, respectively.

### 3.2 Exact Probabilistic Frequent Algorithms

In this subsection, we summarize two existing representative probabilistic frequent itemset mining algorithms: $DP$ (Dynamic Programming-based Apriori algorithm) and $DC$ (Divide-and-Conquer-based Apriori Algorithm). The exact probabilistic frequent itemset mining algorithms first calculate or estimate the frequent probability of each itemset. Then, only for itemsets whose frequent probabilities are larger than the given probability threshold are returned together with their exact frequent probabilities. Because computing the frequent probability is more complicated than calculat-ing the expected support, a quick estimation about whether an itemset is a probabilistic frequent itemset can improve the efficiency of the algorithms. Therefore, a probability tail inequality-based pruning technique, Chernoff bound-based pruning technique, becomes a key tool to improve the efficiency of probabilistic frequent itemset mining algorithms.

**Table 3: Expected Support-based Algorithms**

| Methods | Search Strategy | Data Structure |
|---|---|---|
| UApriori | Breadth-first Search | None |
| UFPgrowth | Depth-first Search | UFP-tree |
| UH-Mine | Depth-first Search | UH-Struct |

### 3.2.1 Dynamic Programming-based Algorithms

Under the definition of the probabilistic frequent itemset, it is critical to compute the frequent probability of an itemset efficiently. [9] is the first work proposing the concept of frequent probability of an itemset and designing a dynamic programming-based algorithm to compute the frequent probability. For the sake of the following discussion, we define that $Pr_{\geq 0,j}(X)$ denotes the probability that itemset $X$ appears at least $i$ times among the first $j$ transactions in the given uncertain database. $Pr(X \subseteq T_j)$ is the probability of itemset $X$ appears in the $j$-th transaction $T_j$. $N$ is the number of transactions in the uncertain database. Therefore, the recursive relationship is defined as follows: $Pr_{\geq i,j}(X) =$

$$Pr_{\geq i,j}(X) \times Pr(X \subseteq T_j) + Pr_{\geq i,j-1}(X) \times (1 - Pr(X \subseteq T_j))$$

Boundary Case: $\begin{cases} Pr_{\geq 0,j}(X) = 1 & 0 \leq j \leq N \\ Pr_{\geq i,j}(X) = 0 & i > j \end{cases}$

Thus, the frequent probability equals $Pr_{\geq min\_sup,N}(X)$. According to the dynamic programming method, $DP$ algorithm uses the Apriori framework to find all probabilistic frequent itemsets.

Based on the definition of the probabilistic frequent itemset, the support of an itemset follows the Poisson Binomial distribution, from which we can deduce that the frequent probability actually equals that one subtracts the probability computed from the corresponding cumulative distribution function (CDF) of the support. Moreover, different from UApriori, $DP$ algorithm computes the frequent probability instead of the expected support for each itemset. The time complexity of the dynamic programming computation for each itemset is $O(N^2 \times min\_sup)$.

### 3.2.2 Divide-and-Conquer-based Algorithms

Besides the dynamic programming-based algorithm, another divide-and-conquer-based algorithm was proposed to compute the frequent probability [28]. Unlike $DP$ algorithm, DC divides an uncertain database, $UDB$, into two sub database: $UDB1$ and $UDB2$. Then, in two sub-databases, the algorithm recursively calls itself to divide the database until only one transaction left. The algorithm stops to record the probability distribution of the support of the itemset in that transaction. Finally, through the conquering part, the complete probability distribution of the itemset support is obtained when the algorithm terminates.

If $DC$ only involves the above divide-and-conquer process, its time complexity of calculating the frequent probability of an itemset is $O(N^2)$ where N is the number of transaction in the uncertain database. However, in the conquering part, $DC$ algorithm can use the Fast Fourier Transform (FFT) method to speed up the efficiency. Thus, the final time complexity of DC algorithm is $O(NlogN)$. In most practical cases, $DC$ algorithm outperforms $DP$ algorithm according to the experimental comparisons reported in Section 4.

### 3.2.3 Effect of the Chernoff Bound-based Pruning

Both the dynamic programming-based and divide-and-conquer-based methods aim to calculate the exact frequent probability for an itemset. However, the computation of the frequent probability is redundant if an itemset is not a probabilistic frequent itemset. Thus, for efficiency improvement, it is a key problem to address how to filter out unpromising probabilistic infrequent itemsets as early as possible. Because the support of an itemset follows Poisson Binomial distribution, Chernoff bound [16] is a well-known tight upper bound of the frequent probability. The Chernoff bound-based pruning is shown as follows.

Table 4: Comparison of Complexity about Determining the Frequent Probability of An Itemset

| Methods | Complexity | Accuracy |
|---|---|---|
| **DP** | $O(N^2 \times min\_sup)$ | Exact |
| **DC** | $O(NlogN)$ | Exact |
| **Chernoff** | $O(N)$ | False Positive |

LEMMA 1. *(Chernoff Bound-based Pruning [28]) Given an uncertain transaction database $UDB$, an itemset $X$, a minimum support threshold $min\_sup$, a probabilistic frequent threshold $pft$, the expected support of $X$, $\mu=esup(X)$, an itemset $X$ is a probabilistic infrequent itemset if,*

$$\begin{cases} 2^{-\delta\mu} < pft & \delta > 2e - 1 \\ e^{-\frac{\delta^2\mu}{4}} < pft & 0 < \delta < 2e - 1 \end{cases}$$

*where $\delta = (min\_sup - \mu - 1)/\mu$ and $N$ is the number of transactions in $UDB$.*

The Chernoff bound can be computed easily as long as the expected support is given. The time complexity of computing the Chernoff bound is $O(N)$ where $N$ is the number of transactions.

**Time Complexity and Accuracy Analysis:** The time complexity and the accuracy of different methods calculating or estimating the frequent probability of an itemset are shown in Table 4. We can find that, it is possible that $DP$ algorithm is faster than DC algorithm if $O(N^2 \times min\_sup) > O(NlogN)$. The Chernoff bound-based pruning spends $O(N)$ to test whether an itemset is not a probabilistic frequent itemset and hence it is the fastest. In addition, with respect to the accuracy, itemsets must be probabilistic frequent itemsets if they can pass the test of $DP$ and $DC$. However, for Chernoff bound-based pruning, there may exist a few false positive results because the Chernoff bound is only an upper bound of the frequent probability.

## 3.3 Approximate Probabilistic Frequent Algorithms

In this subsection, we focus on three approximate probabilistic frequent algorithms. Because the support of an itemset is considered as a random variable following Poisson Binomial distribution under both of definitions, the random variable, i.e., the support of an itemset, can be approximated by the Poisson distribution and the Normal distribution effectively when uncertain databases are large enough. Moreover, for random variables following Poisson distribution and Normal distribution, we can efficiently calculate their probabilities if the expectations and the variances of these random variables are known. Therefore, approximate probabilistic frequent algorithms have the same efficiency of expected support-based algorithms and also guarantee to return frequent probabilities of all probabilistic frequent itemsets with high confidence.

### 3.3.1 Poisson Distribution-based UApriori

In [31], the authors proposed the Poisson distribution-based approximate probabilistic frequent itemset mining algorithm, called PDUApriori. Since we know that the support of an itemset follows Poisson Binomial distribution that can be approximated by the Poisson distribution [12], the frequent probability of an itemset can be rewritten by the cumulative distribution function (CDF) of the Poisson distribution as follows.

$$Pr(X) \approx 1 - e^{-\lambda} \sum_{i=0}^{N \times min\_sup} \frac{\lambda^i}{i!}$$

where $\lambda$ equals the expected support in the above formula since the parameter $\lambda$ in the Poisson distribution is the expectation. PDUApriori algorithm is implemented as follows. Firstly, based on the given probabilistic frequent threshold $pft$, the algorithm computes the corresponding expected support $\lambda$. Then, the algorithm treats $\lambda$ as the minimum expected support and runs the UApriori algorithm to find all the expected support-based frequent itemsets as all the probabilistic frequent itemsets.

PDUApriori utilizes a sound property of the Poisson distribution, namely the fact that parameter $\lambda$ is the expectation and variance of the random variable following the Poisson distribution. Because the cumulative distribution function (CDF) of Poisson distribution is monotonic with respect to $\lambda$, PDUApriori computes the corresponding $\lambda$ of the given $pft$ and calls UApriori to find the results. However, this algorithm only approximately determines whether an itemset is probabilistic frequent itemset, and it cannot return the frequent probability values.

### 3.3.2 Normal Distribution-based UApriori

The Normal distribution-based approximate probabilistic frequent itemset mining algorithm, NDUApriori, was proposed in [10]. According to the Lyapunov Central Limit Theory, Poisson Binomial distribution converges to the Normal Distribution with high probability [25]. Thus, the frequent probability of an itemset can be rewritten by the standard normal distribution formula in the following formula.

$$Pr(X) \approx \Phi(\frac{N \times min\_sup - 0.5 - esup(X)}{\sqrt{Var(X)}})$$

where $\Phi(.)$ is the cumulative distribution function of standard Normal distribution, and $Var(X)$ is the variance of the support of $X$. NDUApriori algorithm employs the Apriori framework and uses the cumulative distribution function of standard Normal distribution to calculate the frequent probability.

Different from PDUApriori, NDUApriori algorithm can return frequent probabilities for all probabilistic frequent itemsets. However, it is impractical to apply NDUApriori on very large sparse uncertain databases since it employs the Apriori framework.

### 3.3.3 Normal Distribution-based UH-Mine

According to the discussion in Section 3.3.1, we can conclude that the UH-mine usually outperforms other expected support-based algorithms in sparse uncertain databases. Moreover, the Normal distribution-based approximation algorithm can acquire the high quality approximate frequent probability. Due to the merits of both the UH-mine and

Table 5: Comparison of Approximate Probabilistic Algorithms

| Methods | Framework | Approximation Methods |
|---|---|---|
| **PDUApriori** | UApriori | Poisson Approximation |
| **NDUApriori** | UApriori | Normal Approximation |
| **NDUH-Mine** | UH-Mine | Normal Approximation |

Normal distribution approximation, we propose a novel algorithm, $NDUH - Mine$ which integrates the framework of UH-Mine and the Normal distribution approximation in order to achieve a win-win partnership in sparse uncertain databases. Compared to UH-Mine, we calculate the variance of each itemset when UH-Mine obtains the expected support of each itemset. In Section 4, we can observe that NDUH-Mine has a better performance than NDUApriori on large sparse uncertain data, which confirms our goal.

Therefore, the Normal distribution-based approximation algorithms build a bridge between the expected support-based frequent itemsets and the probabilistic frequent itemsets. In particular, existing efficient expected support-based mining algorithms can directly be reused in the problem of mining probabilistic frequent itemsets and keep their intrinsic properties. In other words, under the definition of mining probabilistic frequent itemsets, NDUApriori is the fastest algorithm in large enough dense uncertain database, while NDUH-Mine requires reasonable memory space and scales well to very large sparse uncertain databases.

**Comparison of Algorithm Framework and Approximation Methods:** Different from the exact probabilistic frequent itemset mining algorithms, the computational complexities of computing the frequent probability of each itemset for different approximate probabilistic frequent algorithms are the same, $O(N)$ where $N$ is the number of transactions of the given uncertain database. Thus, we mainly compare the different algorithm frameworks and approximation approaches for the three approximate probabilistic frequent itemset mining algorithms in Table 5.

## 4. EXPERIMENTS
### 4.1 Experimental Settings

In this subsection, we introduce the experimental environment, the implementations, and the evaluation methods.

Firstly, in order to conduct a fair comparison, we build a common implementation framework which provides common data structures and subroutines for implementing all the algorithms. All the experiments are performed on an Intel(R) Core(TM) i7 3.40GHz PC with 4GB main memory, running on Microsoft Windows 7. Moreover, all algorithms were implemented and compiled using Microsoft's Visual C++ 2010.

For each algorithm, we use the existing robust implementation for our comparison. According to the discussion in Section 3, we separate comparisons into the three categories: expected support-based algorithms; exact probabilistic frequent algorithms; approximation probabilistic frequent algorithms. For expected support algorithms, we use the version in [17, 18] to implement the UApriori algorithm which employed decremental pruning and hashing techniques to speed up the mining process. The implementation based on [22] is used to test UFP-growth algorithm. We do not use the UCFP-tree implementation [4] since there is no obvious optimization between UFP-growth algorithm and UCFP-tree algorithm in terms of the running time and the memory cost. The UH-Mine algorithm is implemented based on the version in [4]. For four exact probabilistic frequent algorithms, DPNB (Dynamic Programming-based Algorithm with No Bound) algorithm that does not include the Chernoff bound-based pruning technique is implemented based on the version in [9]. Correspondingly, DCNB (Divide-and-Conquer-based Algorithm No Bound) algorithm is modified based on the version in [28]. However, what is different is in our algorithm, each item has their own probability, while in [28], all items in a transaction share the same appearance probability. Moreover, DPB (Dynamic Programming-based Algorithm with Bound) algorithm [9] and DCB (Divide-and-Conquer-based Algorithm with Bound) algorithm [28] represent the corresponding algorithms of DPNB and DCNB, but include the Chernoff bound-based pruning, respectively. For three approximation mining algorithms, the implementation of PDUApriori is based on [31] and integrates all optimized pruning techniques. PDUApriori [10] and PDUH-Mine are implemented based on the frameworks of UApriori and UH-Mine, respectively. Hashing function is used in the two algorithms to compute the cumulative distribution function of Standard Normal Distribution efficiently.

Based on the experimental comparisons of existing researches, we choose five classical deterministic benchmarks from FIMI repository [1], and assign a probability generated from Gaussian distribution to each item. Assigning probability to deterministic database to generate meaningful uncertain test data is widely accepted by the current community [4, 9, 10, 11, 17, 18, 22, 28, 30, 31]. Five datasets includes two dense datasets, Connect and Accident, two sparse datasets, Kosarak and Gazelle, and a very large synthetic dataset T25I15D320k which was used for testing the scalability of uncertain frequent itemset mining algorithms [4]. The characteristics of above datasets are shown in Table 6. In addition, to verify the influence of uncertainty, we also test another probability distribution, Zipf distribution, instead of Gaussian distribution. Among the cases that datasets following the Gaussian distribution, we further categorize four scenarios. The first scenario is that a dense dataset with high mean and low variance, namely Connect with the mean (0.95) and the variance (0.05). The second scenario is that a dense dataset with low mean and high variance, namely Accident with the mean (0.5) and the variance (0.5). The third scenario is that a sparse dataset with high mean and low variance, namely Gazelle with the mean (0.95) and the variance (0.05). The fourth scenario is that a sparse dataset with high mean and low variance, namely Kosarak with the mean (0.5) and the variance (0.5). Moreover, in the case that dataset following the Zipf distribution, the only one scenario that a dense dataset following the Zipf distribution varying the skew from 0.8 to 2 is tested because the sparse datasets followed Zipf distribution only have a very small size of frequent itemsets, thus we did not get any meaningful results from these datasets.

The probability parameters and their default values of each datasets are shown in Table 7. For all the tests, we perform 10 runs per experiment and report the averages. In addition, we do not report the running time over 1 hour.

## 4.2 Expected Support-based Algorithms

In this section, we compare three expected support-based algorithms, UApriori, UFP-growth, and UH-Mine. Firstly,

Table 6: Characteristics of Datasets

| Dataset | # of Trans. | # of Items | Ave. Len. | Density |
|---|---|---|---|---|
| Connect | 67557 | 129 | 43 | 0.33 |
| Accident | 340183 | 468 | 33.8 | 0.072 |
| Kosarak | 990002 | 41270 | 8.1 | 0.00019 |
| Gazelle | 59601 | 498 | 2.5 | 0.005 |
| T25I15D320k | 320,000 | 994 | 25 | 0.025 |

Table 7: Default Parameters of Datasets

| Dataset | Mean | Var. | min_sup | pft |
|---|---|---|---|---|
| Connect | 0.95 | 0.05 | 0.5 | 0.9 |
| Accident | 0.5 | 0.5 | 0.5 | 0.9 |
| Kosarak | 0.5 | 0.5 | 0.0005 | 0.9 |
| Gazelle | 0.95 | 0.05 | 0.025 | 0.9 |
| T25I15D320k | 0.9 | 0.1 | 0.1 | 0.9 |

we report the running time and the memory cost in two dense datasets and two sparse datasets. Secondly, we present the scalability of the three algorithms. Finally, we study the influence of the skew in the Zipf distribution.

**Running Time.** Figures 4(a) - 4(d) show the running time of expected support-based algorithms w.r.t $min\_esup$ in Connect, Accident, Kosarak, and Gazelle datasets. When $min\_esup$ decreases, we observe that the running time of all the algorithms goes up. Moreover, UFP-growth is always the slowest in the above results. UApriori is faster than UH-Mine in Figures 4(a) and 4(b), on the other hand, UH-Mine is faster than UApriori in Figures 4(c) and 4(d).

It is reasonable because UApriori outperforms other algorithms under the conditions that uncertain dataset is dense and $min\_esup$ is high enough. These conditions cause the search space of mining algorithm relatively small. Under this case, the breath-first-search-based algorithms are faster than the depth-first-search-based algorithms. Thus, UApriori outperforms the other two depth-first-search-based algorithms in Figure 4(a) and 4(b). Otherwise, the depth-first-search-based algorithm, UH-Mine, is better, which is proved by Figure 4(c) and 4(d). However, even UFP-growth using depth-first-search strategy, it does not perform well because UFP-growth spends too much time on recursively constructing many redundant conditional subtrees with limited shared paths. In addition, another interesting observation is that slopes of curves in Figure 4(a) are larger than those in 4(c) even if the size of Connect is smaller than that of Kosarak. This result makes sense because the slope of the curve depends on the density of a dataset.

**Memory Cost.** According to Figures 4(e) - 4(f), UFP-growth spends memory the most among all the three algorithms. Similar to the conclusion given in the above analysis of Running Time, UApriori is superior to UH-Mine if and only if the uncertain dataset is dense and $min\_esup$ is high enough, otherwise, UH-Mine is the winner.

UApriori uses less memory when $min\_esup$ is high and dataset is dense because there are small number of frequent itemsets generated which results in small search space. However, as $min\_esup$ decreasing and dataset becoming sparse, UApriori has to require much more memory to store redundant infrequent candidates. Thus, the memory usage trend of UApriori changes sharply with decreasing $min\_esup$. For UH-Mine, the main memory cost is used to initialize its UH-Struct. However, with the increase of $min\_esup$, UH-Struct only spends the limited memory cost on building the head tables for different prefixes. Therefore, the memory usage of UH-Mine increases smoothly. Moreover, similar to the discussion of Running Time, UFP-growth is the most memory consuming one among three algorithms.

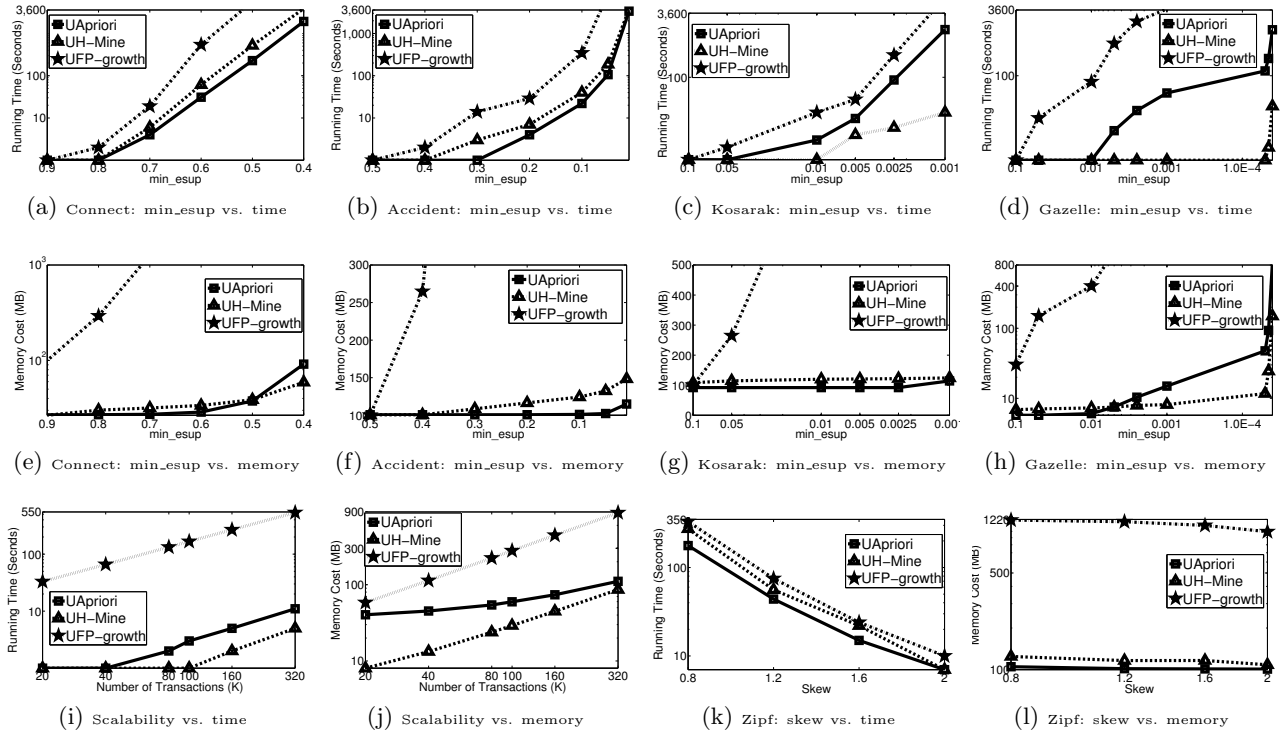

Figure 4: Performance of Expected Support-based Frequent Algorithms

**Scalability.** We further analyze the scalability of three expected support-based algorithms. In Figure 4(i), varying the number of transactions in the dataset from 20k to 320k, we observe that the running time is linear. With the increase of the size of dataset, the time of UApriori is close to that of UH-mine. This is reasonable because all the items in T25I15D30k have similar distributions. Therefore, with the increase of transactions, the running time of algorithms increase linearly. Figure 4(j) reports the memory usages of three algorithms which demonstrate the linearity in terms of the number of transactions. Moreover, we can find that the memory usage increase of UApriori is more steady than that of two other algorithms. This is because UApriori needs not to build a special data structure to store the uncertain database. However, the two other algorithms have to spend the extra memory cost for storing their data structures. Therefore, the curve of UApriori is more steady.

**Effect of the Zipf distribution.** For verifying the influence of uncertainty under different distributions, Figures 4(k) and 4(l) show the running time and the memory cost of three algorithms in terms of the skew parameter of Zipf distribution. We can observe that the running time and the memory cost decrease with the increase of the skew parameter. Due to the property of Zipf distribution, more items are assigned the zero probability with the increase of the skew parameter, which results in fewer frequent itemsets. Specifically, when the skew parameter increases, we can observe that UH-Mine outperforms UApriori gradually.

**Conclusions.** To sum up, under the definition of expected support-based frequent itemset, there is no clear winner among current proposed mining algorithms. In the condition of dense datasets and higher $min\_esup$, UApriori spends the least time and memory. Otherwise, UH-Mine is the winner.

Moreover, UFP-growth is often the slowest algorithm and spends the largest memory cost since UFP-growth has only limited shared paths so that it has to spend too much time and memory on redundant recursive computation.

Finally, the influence of the Zipf distribution is similar to that of a very sparse dataset. Under the Zipf distribution, UH-Mine algorithm usually performs very well.

### 4.3 Exact Probabilistic Frequent Algorithms

In this section, we compare four probabilistic frequent algorithms: DPNB, DCNB, DPB and DCB. Firstly, we show the running time and the memory cost in terms of changing $min\_sup$. Then, we present the influence of $pft$ on the running time and the memory cost. Moreover, the scalability of the three algorithms is studied. Finally, we report the influence of the skew in the Zipf distribution as well.

**Effect of min_sup.** Figures 5(a) and 5(c) show the running time of four competitive algorithms w.r.t. $min\_sup$ in Accident and Kosarak datasets, respectively. With the Chernoff-bound-based pruning, we can see that DCB is always faster than DPB. However, without the Chernoff-bound-based pruning, we can find that DCNB is always faster than DPNB. This is reasonable because the time complexity of computing the frequent probability of each itemset in divide-and-conquer-based algorithms is $O(NlogN)$, which is more efficient than that of dynamic programming-based algorithms, $O(N^2 \times min\_sup)$. Comparing the same type of algorithms, we can find that DCB is faster than DCNB and DPB is faster than DPNB. These results show that most infrequent itemsets can be filtered by the Chernoff bound-based pruning quickly. Moreover, we also observe that DPB is faster than DCNB, this is because there are only a smal-

1657

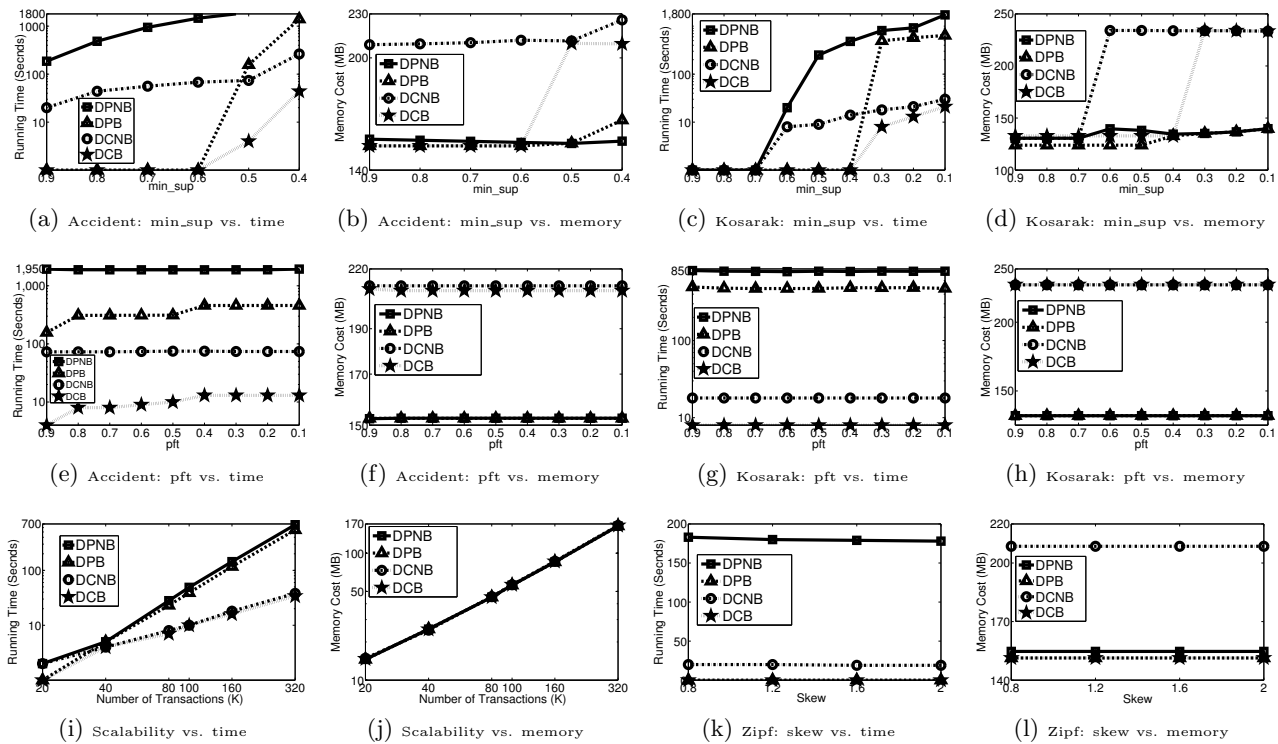

**Figure 5: Performance of Exact Probabilistic Frequent Algorithms**

l number of frequent itemsets that need to compute their frequent probabilities when $min\_sup$ is high, most of the infrequent itemsets are already pruned by the Chernoff bound.

In addition, according to Figures 5(b) and 5(d), this is very clear that DPB and DPNB require less memory than DCB and DCNB. It is reasonable because both DCB and DCNB trade off the memory for the efficiency based on their the divide-and-conquer strategy. In addition, we can observe that the memory usage trend of DCNB changes sharply with decreasing $min\_sup$ because there are a few frequent itemsets when $min\_sup$ is high and most of infrequent itemsets are filtered out by the Chernoff bound-based pruning. In particular, we can find that similar observations w.r.t $min\_sup$ are shown in both the dense and the sparse datasets, which indicate that the density of the databases is not the key factor affecting the running time and the memory usage of exact probabilistic frequent algorithms.

**Effect of pft.** Figures 5(e) and 5(g) report the running time w.r.t. $pft$. We can find that DCB is still the fastest algorithm and DPNB is the slowest one. Different from the results w.r.t. $min\_sup$, DCNB is always faster than DPB when $pft$ varies. Additionally, Figures 5(f) and 5(h) show the memory cost w.r.t. $pft$. The memory usages of both DPB and DPNB are always significantly smaller than those of both DCB and DCNB. Furthermore, we find that, by varying $pft$, the changing trends of the runing time and the memory cost are quite stable. Thus, $pft$ does not have significant impact to the running time and the memory of the four mining algorithms. This is reasonable because most frequent probabilities of frequent itemsets are one and it is also further explained in the next subsection.

**Scalability.** Similar to the scalability analysis in Section 4.2, we still use the T25I15D320k dataset to test the scalability of four exact probabilistic frequent itemset mining algorithm. In Figures 5(i), we can find that the trends of running time of all algorithms are linear with the increase of the number of transactions. In particular, the trends of both DC and DCNB are more smooth than those of DP and DPNB because the time complexities of computing frequent probability for DC and DCNB are both $O(NlogN)$ and better than the time complexities of DP and DPNB. In Figures 5(j), we can observe that the memory cost of four algorithms linearly varies w.r.t. the number of transactions.

**Effect of the Zipf distribution.** Figures 5(k) and 5(l) show the running time and the memory cost of four exact probabilistic frequent mining algorithms in terms of the skew parameter of Zipf distribution. We can observe that the running time and the memory cost decrease with the increase of the skew parameter. We can find that, through varying the skew parameter, the changing trends of the runing time and the memory cost are quite stable. Therefore, the skew parameter of Zipf distribution does not have significant impact to the running time and the memory cost.

**Conclusions.** First of all, among exact probabilistic frequent itemsets mining algorithms, DCB algorithm is the fastest algorithm in most cases. However, compared to DPB, it has to spend more memory for the divide-and-conquer processing.

In addition, the Chernoff bound-based pruning is the most important tool to speed up exact probabilistic frequent itemset mining algorithms. Based on computational analysis, the computing Chernoff bound of each itemset is only $O(N)$. However, DC and DP algorithms have to spend $O(NlogN)$ and $O(N^2 \times min\_sup)$ to calculate the exact frequent probability for each an itemset, respectively. Therefore, it is clear that Chernoff bound-based pruning can reduce the running time if it can filter out some infrequent itemsets.

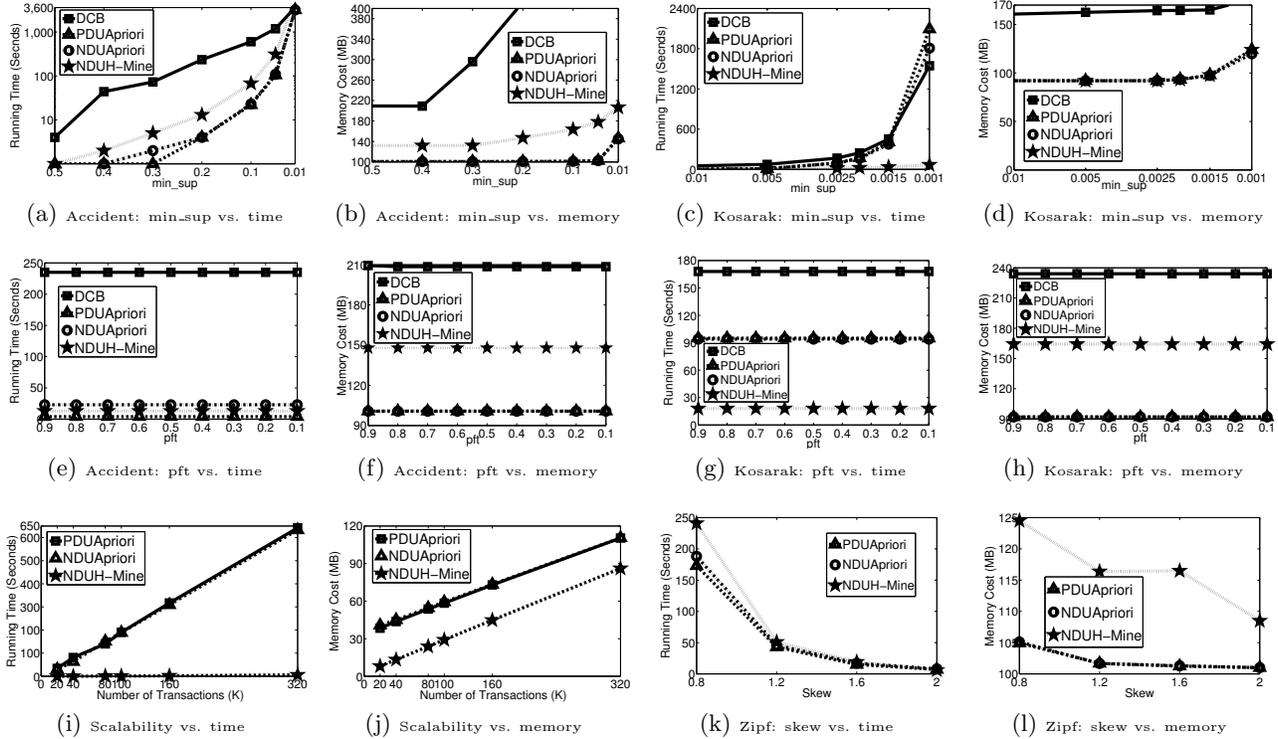

Figure 6: Performance of Approximation Probabilistic Frequent Algorithms

## 4.4 Approximate Probabilistic Frequent Algorithms

In this section, we mainly compare three approximation probabilistic frequent algorithms, PDUApriori, NDUApriori, and NDUH-Mine, and an exact probabilistic frequent algorithm, DCB. Firstly, we report the running time and the memory cost in terms of $min\_sup$. Then, we present the running time and the memory cost when $pft$ is changed. In addition, we test the precsion and the recall to evaluate the approximation quality. Finally, we report the scalability as well.

**Effect of min_sup.** First of all, we test the running time and memory cost of four algorithms w.r.t. the minimum support, $min\_sup$ shown in Figures 6(a) - 6(d). In Figure 6(a), both PDUApriori and NDUApriori are faster than the other two. In Figure 6(c), NDUH-Mine is the fastest. Moreover, DCB is the slowest algorithm among the four algorithms since it offers exact answers. This is reasonable because PDUApriori and NDUApriori are based on the UApriori framework which performs best under the conditions that uncertain dataset is dense and $min\_sup$ is enough high. Otherwise, NDUH-Mine is the best.

In addition, in Figures 6(b) and 6(d), it is very clear that P-DUApriori, NDUApriori, and NDUH-Mine require less memory than DCB. This is reasonable because DCB uses the divide-and-conquer strategy to obtain the exact results. In Figure 6(b), both PDUApriori and NDUApriori require less memory because the dataset is dense. In Figure 6(d), NDUH-Mine spends less memory since this dataset is sparse.

**Effect of pft.** Figure 6(e) reports the time in terms of varying $pft$. We can see that both of PDUApriori and ND-UApriori are still the fastest algorithms in Accident dataset. However, in Figure 6(g), NDUH-Mine is the fastest algorithm. The results also confirm that the density of databases is the most important factor for the approximate algorithm efficiency again. However, Figure 6(f) shows that the memory cost of all four algorithms is steady. Similar results are also shown in Figure 6(h). Hence, varying $pft$ almost does not influence the memory cost of algorithms.

**Precision and Recall.** Besides offering efficient running time and effective memory cost, the approximation accuracy is a more important target for the approximation probabilistic frequent itemset mining algorithms. We use the precision which equals $\frac{|AR \cap ER|}{|AR|}$ and the recall which equals $\frac{|AR \cap ER|}{|ER|}$ to measure the accuracy of approximation probabilistic frequent mining algorithm. Please note that AR means the result generated from the approximation probabilistic frequent algorithm, and ER is the result generated from the exact probabilistic frequent algorithm. Moreover, we only test the precision and the recall w.r.t varying $min\_sup$ be-cause the influence of $pft$ is far less than the $min\_sup$. Table 8 and Table 9 are shown the precisions and the recalls of two approximation probabilistic frequent algorithms in Accident and Kosarak, respectively. We can find that the precision and the recall are almost 1 in Accident dataset which means there is almost no false positive and false negative. In Kosarak, we also observe that there are a few false positives with decreasing of $min\_sup$. In addition, the Normal distribution-based approximation algorithms can get better approximation effect than the Poisson distribution-based approximation algorithms. This is because the expectation and the variance in the Poisson distribution is the same, which is $\lambda$, but, in fact, the expected support and the variance of an itemset are usually unequal.

Table 8: Accuracy in Accident

| Min_Sup | PDUApriori | | NDUApriori | | NDUH-Mine | |
|---|---|---|---|---|---|---|
| | P | R | P | R | P | R |
| 0.2 | 0.91 | 1 | 0.95 | 1 | 0.95 | 1 |
| 0.3 | 1 | 1 | 1 | 1 | 1 | 1 |
| 0.4 | 1 | 1 | 1 | 1 | 1 | 1 |
| 0.5 | 1 | 1 | 1 | 1 | 1 | 1 |
| 0.6 | 1 | 1 | 1 | 1 | 1 | 1 |

Table 9: Accuracy in Kosarak

| Min_Sup | PDUApriori | | NDUApriori | | NDUH-Mine | |
|---|---|---|---|---|---|---|
| | P | R | P | R | P | R |
| 0.0025 | 0.95 | 1 | 0.95 | 1 | 0.95 | 1 |
| 0.005 | 0.96 | 1 | 0.96 | 1 | 0.96 | 1 |
| 0.01 | 0.98 | 1 | 0.98 | 1 | 0.98 | 1 |
| 0.05 | 1 | 1 | 1 | 1 | 1 | 1 |
| 0.1 | 1 | 1 | 1 | 1 | 1 | 1 |

Table 10: Summary of Eight Representative Frequent Itemset Algorithms over Uncertain Databases

|  | Expected Support-based Alg | | | Exact Prob. Freq. Alg | | Approx. Prob. Freq. Alg | | |
| --- | --- | --- | --- | --- | --- | --- | --- | --- |
|  | UApriori | UH-Mine | UFP-growth | DP | DC | PDUApriori | NDUApriori | NDUH-Mine |
| Time(D) | √($min\_esup$ high) | √($min\_esup$ low) |  |  | √ | √($min\_sup$ high) | √($min\_sup$ high) | √($min\_sup$ low) |
| Time(S) |  | √ |  |  | √ |  |  | √ |
| Memory(D) | √($min\_esup$ high) | √($min\_esup$ low) |  | √ |  | √($min\_sup$ high) | √($min\_sup$ high) | √($min\_sup$ low) |
| Memory(S) |  | √ |  | √ |  |  |  | √ |
| Accuracy | Exact | Exact | Exact | Exact | Exact | Approx. | Approx.(Better) | Approx.(Better) |

**Scalability.** We further analyze the scalability of three approximate probabilistic frequent mining algorithms. In Figure 6(i), varying the number of transactions in the dataset from 20k to 320k, we find that the running time is linear. Figure 6(j) reports the memory cost of three algorithms which show the linearity in terms of the number of transactions. Therefore, NDUH-Mine performs best.

**Effect of the Zipf distribution.** Figures 6(k) and 6(l) show the running time and the memory cost of three approximate algorithms in terms of the skew parameter of Zipf distribution. We can observe that the running time and the memory cost decrease with the increase of the skew parameter. In particular, when the skew parameter increases, we can observe that PDUApriori outperforms NDUApriori and NDUH-Mine gradually.

**Conclusions.** First of all, approximation probabilistic frequent itemset mining algorithms can get high-quality approximation when the uncertain database is large enough due to the requirement of CLT. In our experiments, the datasets usually include more than 50,000 transactions. These approximation algorithms almost have no false positive or false negative. These results are reasonable because the Lyapunov CLT guarantees the approximation quality.

In addition, in terms of the efficiency, approximation probabilistic frequent itemset mining algorithms are much better any existing exact probabilistic frequent itemset mining algorithms. Moreover, Normal distribution-based algorithms usually are faster than the Poisson distribution-based algorithm.

Finally, similar to the case of expected support-based frequent algorithms, NDUApriori is always the fastest algorithm in dense uncertain databases, while NDUH-Mine usually the best algorithm in sparse uncertain databases.

### 4.5 Summary of New Findings

We summarize experimental results under different cases in Table 10 where '√' means the winner in that case. Moreover, 'time(D)' means that the time cost in the dense data set, and 'time(S)' means that the time cost in the sparse data set. The meanings of 'memory(D)' and 'memory(S)' are similar.

- As observed in Table 10, under the definition of expected support-based frequent itemset, UApriori is usually the fastest algorithm with lower memory cost when the database is dense and $min\_sup$ is high. On the contrary, when the database is sparse or $min\_sup$ is low, UH-Mine often outperforms other algorithms in the running time and only spends limited memory cost. However, UFP-growth is almost the slowest algorithm with high memory cost.

- From Table 10, among exact probabilistic frequent itemsets mining algorithms, DC algorithm is the fastest algorithm in most cases. However, it trades off the memory cost for the efficiency because it has to store recursive results for the processing of the divide-and-conquer. In addition, when the condition is satisfied, DP algorithm is faster than DC algorithm.

- Again from Table 10, both PDUApriori and NDUApriori is the winner in the running time and the memory cost when the database is dense and $min\_sup$ is high, otherwise, NDUH-Mine is the winner. The main difference between PDUApriori and NDUApriori is that NDUApriori has better approximation when the database is large enough.

Other than the result described in Table 10, we also find:

- Approximation probabilistic frequent itemset mining algorithms usually get a high-quality approximation effect in most cases. To our surprise, the frequent probabilities of most probabilistic frequent itemsets are often 1 when the uncertain databases are large enough such as the number of transaction is more than 10,000. It is a reasonable result. On the one hand, Lyapunov Central Limit Theory guarantees the high-quality approximation. On the other hand, according to the cumulative distribution function (CDF) of the Poisson distribution, we know that the frequent probability of an itemset can be approximated as $1-e^{-\lambda}\sum_{i=0}^{N\times min\_sup}\frac{\lambda^i}{i!}$ where $\lambda$ is the expected support of this itemset. When an uncertain database is large enough, the expected support of this itemset is usually large if it is a probabilistic frequent itemset. Thus, as a consequence, the frequent probability of this itemset equals 1.

- Approximation probabilistic frequent itemset mining algorithms usually far outperform any existing exact probabilistic frequent itemset mining algorithms in the algorithm efficiency and the memory cost. Therefore, the result under the definition of probabilistic frequent itemset can be obtained by the existing solutions under the definition of expected support-based frequent itemset if we compute the variance of the support of itemsets as well.

- Chernoff bound is an important tool to improve the efficiency of exact probabilistic frequent algorithms because it can filter out the infrequent itemsets quickly.

## 5. CONCLUSIONS

In this paper, we conduct a comprehensive experimental study of all the frequent itemset mining algorithms over uncertain databases. Since there are two definitions of frequent itemsets over uncertain data, most existing researches are categorized into two directions. However, through our exploration, we firstly clarify that there is a close relationship between two different definitions of frequent itemsets over uncertain data. Therefore, we need not use the current solution for the second definition and replace them with efficient existing solution of first definition. Secondly, we provide baseline implementations of eight existing representative algorithms and test their performances under a uniform measurement fairly. Finally, based on extensive experiments over many different benchmarks, we verify several existing inconsistent conclusions and find some new rules in this area.

## 6. ACKNOWLEDGMENTS

This work is supported in part by the Hong Kong RGC GRF Project No.611411, National Grand Fundamental Research 973 Program of China under Grant 2012-CB316200 and 2011-CB302200-G, HP IRP Project 2011, Microsoft Research Asia Grant, MRA11EG05, the National Natural Science Foundation of China (Grant No.61025007, 60933001, 61100024), US NSF grants DBI-0960443, CNS-1115234, and IIS-0914934, and Google Mobile 2014 Program.